\documentclass[11pt]{article}

\usepackage{amssymb, amsmath, amsfonts, amsbsy, amsbsy, latexsym, color, epsfig, graphicx, url}

\newcommand{\bitem}{\begin{itemize}}
\newcommand{\eitem}{\end{itemize}}
\newcommand{\benum}{\begin{enumerate}}
\newcommand{\eenum}{\end{enumerate}}
\newcommand{\beq}{\begin{equation}}
\newcommand{\eeq}{\end{equation}}
\newcommand{\ip}[2]{\langle#1,#2\rangle}

\newcommand{\absip}[2]{| \langle#1,#2\rangle |}
\newcommand{\norm}[1]{\|#1\|}

\newcommand{\argmin}{\mbox{\rm argmin}}

\DeclareMathOperator*{\supp}{supp}

\def\NN{\mathbb{N}}

\def\RR{\mathbb{R}}

\def\cH{{\mathcal{H}}}

\def\cW{{\mathcal{W}}}

\def\cM{{\mathcal{M}}}
\def\cN{{\mathcal{N}}}

\def\cS{{\mathcal{S}}}

\newcommand{\qed}{$\Box$}

\newtheorem{theorem}{Theorem}[section]

\newtheorem{lemma}{Lemma}[section]
\newtheorem{definition}{Definition}[section]

\title{Theory and Applications of Compressed Sensing}
\author{Gitta Kutyniok}

\begin{document}
\maketitle

\begin{abstract}
Compressed sensing is a novel research area, which was introduced in
2006, and since then has already become a key concept in various
areas of applied mathematics, computer science, and electrical
engineering. It surprisingly predicts that high-dimensional signals,
which allow a sparse representation by a suitable basis or, more
generally, a frame, can be recovered from what was previously
considered highly incomplete linear measurements by using efficient
algorithms. This article shall serve as an introduction to and a
survey about compressed sensing.
\end{abstract}

\vspace{.1in} {\bf Key Words.} Dimension reduction. Frames. Greedy
algorithms. Ill-posed inverse problems. $\ell_1$ minimization.
Random matrices. Sparse approximation. Sparse recovery.\vspace{.1in}

\vspace{.1in} {\bf Acknowledgements.}
The author is grateful to the reviewers for many helpful suggestions which
improved the presentation of the paper. She would also like to thank Emmanuel Cand\`es, David Donoho,
Michael Elad, and Yonina Eldar for various discussions on related topics, and Sadegh Jokar
for producing Figure 3.
The author acknowledges support by the Einstein
Foundation Berlin, by Deutsche Forschungsgemeinschaft (DFG) Grants SPP-1324 KU 1446/13 and KU 1446/14,
and by the DFG Research Center {\sc Matheon} ``Mathematics for key technologies'' in Berlin.

\vspace{.1in}

\section{Introduction}

The area of compressed sensing was initiated in 2006 by two groundbreaking
papers, namely \cite{Don06c} by Donoho and \cite{CRT06} by
Cand\`{e}s, Romberg, and Tao. Nowadays, after only 6 years, an
abundance of theoretical aspects of compressed sensing are
explored in more than 1000 articles. Moreover, this methodology is
to date extensively utilized by applied mathematicians, computer
scientists, and engineers for a variety of applications in
astronomy, biology, medicine, radar, and seismology,  to name a few.

The key idea of compressed sensing is to recover a sparse signal
from very few non-adaptive, linear measurements by convex
optimization. Taking a different viewpoint, it concerns the exact
recovery of a high-dimensional sparse vector after a dimension
reduction step. From a yet another standpoint, we can regard
the problem as computing a sparse coefficient vector for a signal
with respect to an overcomplete system. The theoretical foundation
of compressed sensing has links with and also explores methodologies from
various other fields such as, for example, applied harmonic
analysis, frame theory, geometric functional analysis, numerical
linear algebra, optimization theory, and random matrix theory.

It is interesting to notice that this development -- the problem of
sparse recovery -- can in fact be traced back to earlier papers from
the 90s such as \cite{DS89} and later the prominent papers by
Donoho and Huo \cite{DH01} and Donoho and Elad \cite{DE03}. When the
previously mentioned two fundamental papers introducing compressed sensing were published, the term `compressed sensing' was initially
utilized for random sensing matrices, since those allow for a
minimal number of non-adaptive, linear measurements. Nowadays, the
terminology `compressed sensing' is more and more often used
interchangeably with `sparse recovery' in general, which is a
viewpoint we will also take in this survey paper.

\subsection{The Compressed Sensing Problem}

To state the problem mathematically precisely, let now $x =
(x_i)_{i=1}^n \in \RR^n$ be our signal of interest. As prior
information, we either assume that $x$ itself is {\em sparse}, i.e.,
it has very few non-zero coefficients in the sense that
\[
\|x\|_0 := \#\{i : x_i \neq 0\}
\]
is small, or that there exists an orthonormal basis or a
frame\footnote{Recall that a {\em frame} for a Hilbert space $\cH$
is a system $(\varphi_i)_{i\in I}$ in $\cH$, for which there exist
{\em frame bounds} $0 < A \le B < \infty$ such that $A \|x\|_2^2 \le
\sum_{i\in I} |\ip{x}{\varphi_i}|^2 \le B \|x\|_2^2$ for all $x \in \cH$. A
{\em tight frame} allows $A=B$. If $A=B=1$ can be chosen,
$(\varphi_i)_{i\in I}$ forms a {\em Parseval frame}. For further
information, we refer to \cite{CK12}.} $\Phi$ such that $x = \Phi c$
with $c$ being sparse. For this, we let $\Phi$ be the matrix with
the elements of the orthonormal basis or the frame as column
vectors. In fact, a frame typically provides more flexibility than an
orthonormal basis due to its redundancy and hence leads to improved
sparsifying properties, hence in this setting customarily frames
are more often employed than orthonormal bases. Sometimes the notion
of sparsity is weakened, which we for now -- before we will make this
precise in Section \ref{sec:signalmodels} -- will refer to as {\em
approximately sparse}. Further, let $A$ be an $m \times n$ matrix,
which is typically called {\em sensing matrix} or {\em measurement
matrix}. Throughout we will always assume that $m < n$ and that
$A$ does not possess any zero columns, even if not explicitly mentioned.

Then the {\em Compressed Sensing Problem} can be formulated as
follows: Recover $x$ from knowledge of
\[
y = Ax,
\]
or recover $c$ from knowledge of
\[
y = A \Phi c.
\]
In both cases, we face an underdetermined linear system of equations
with sparsity as prior information about the vector to be recovered
-- we do {\em not} however know the support, since then the solution could be
trivially obtained.

This leads us to the following questions:\\[-3ex]
\bitem
\item What are suitable signal and sparsity models?\\[-4ex]
\item How, when, and with how much accuracy can the signal be algorithmically recovered?\\[-4ex]
\item What are suitable sensing matrices?
\eitem
In this section, we will discuss these questions briefly to build up
intuition for the subsequent sections.

\subsection{Sparsity: A Reasonable Assumption?}

As a first consideration, one might question whether sparsity is
indeed a reasonable assumption. Due to the complexity of real data
certainly only a heuristic answer is possible.

\begin{figure}[h]
\centering
\includegraphics[width=6cm]{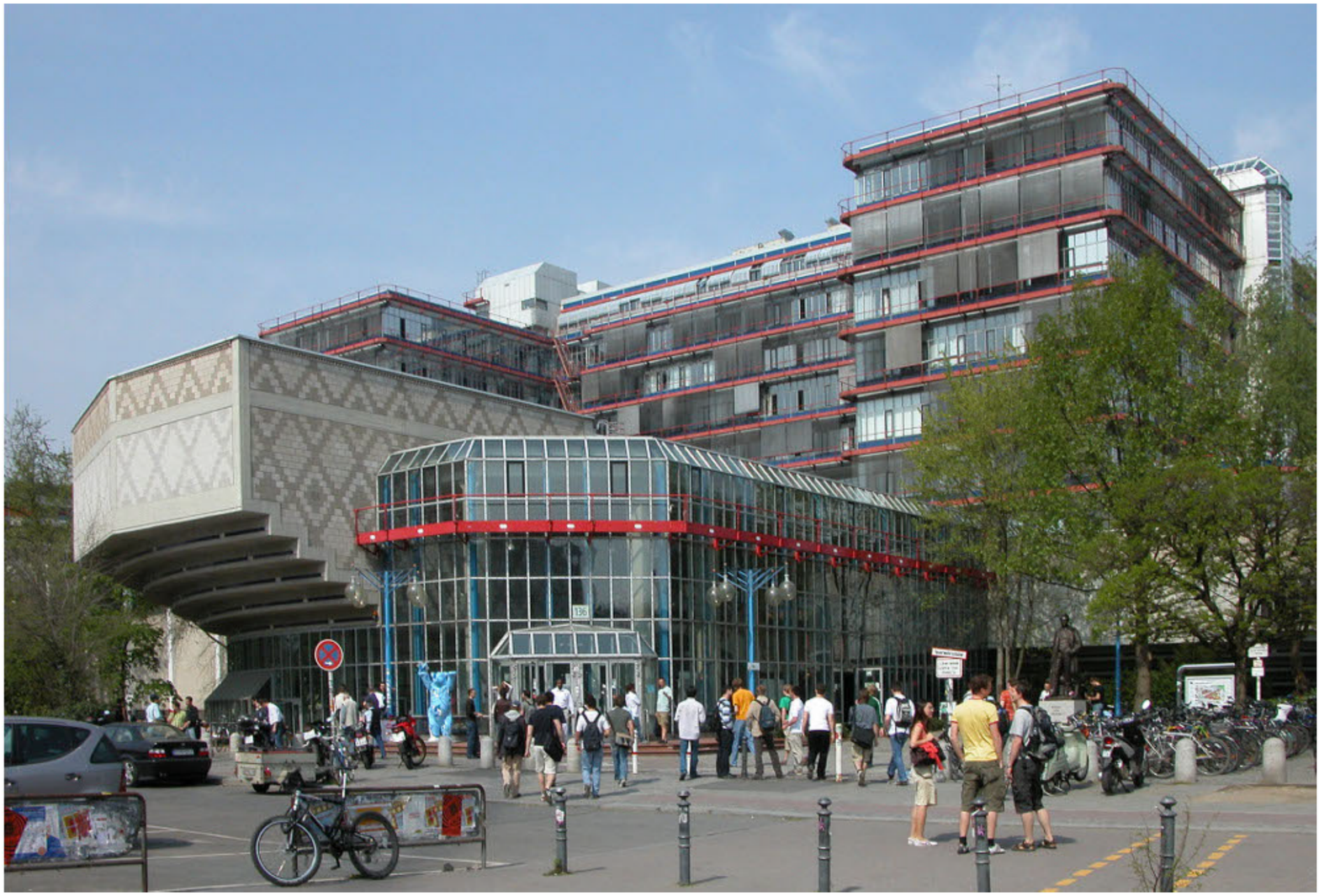}
\hspace*{0.5cm}
\includegraphics[width=6cm]{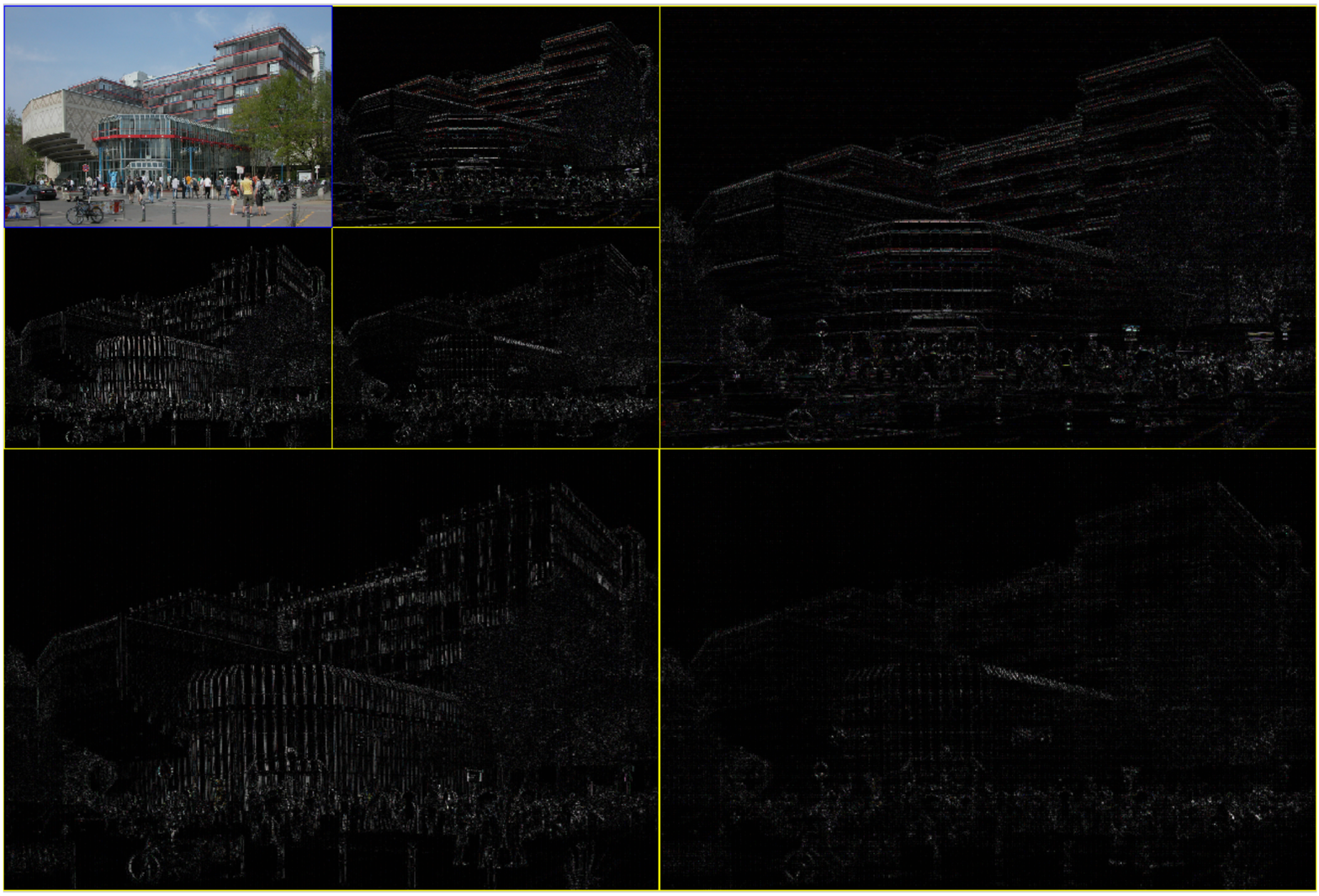}
\put(-280,-15){(a)} \put(-90,-15){(b)} \caption{(a) Mathematics
building of TU Berlin (Photo by TU-Pressestelle); (b) Wavelet
decomposition} \label{fig:sparsity}
\end{figure}

If a natural image is taken, it is well known that wavelets
typically provide sparse approximations. This is illustrated in
Figure \ref{fig:sparsity}, which shows a wavelet decomposition
\cite{Mal98} of an exemplary image. It can clearly be seen that most
coefficients are small in absolute value, indicated by a darker
color.

Depending on the signal, a variety of representation systems which
can be used to provide sparse approximations is available and is
constantly expanded. In fact, it was recently shown that wavelet
systems do not provide optimally sparse approximations in a regularity
setting which appears to be suitable for most natural images,
but the novel system of shearlets does
\cite{KL10a,KL12}. Hence, assuming some prior knowledge of the signal
to be sensed or compressed, typically suitable, well-analyzed
representation systems are already at hand. If this is not the case,
more data sensitive methods such as dictionary learning algorithms (see, for instance, \cite{AEB06}),
in which a suitable representation system is computed for a given set
of test signals, are available.

Depending on the application at hand, often $x$ is already sparse
itself. Think, for instance, of digital communication, when a cell
phone network with $n$ antennas and $m$ users needs to be modelled.
Or consider genomics, when in a test study $m$ genes shall be analyzed with
$n$ patients taking part in the study. In the first scenario, very
few of the users have an ongoing call at a specific time; in the
second scenario, very few of the genes are actually active. Thus,
$x$ being sparse itself is also a very natural assumption.

In the compressed sensing literature, most results indeed assume
that $x$ itself is sparse, and the problem $y=Ax$ is considered.
Very few articles study the problem of incorporating a sparsifying
orthonormal basis or frame; we mention specifically \cite{RSV08,CENP11}. In
this paper, we will also assume throughout that $x$ is already a
sparse vector. It should be emphasized that `exact' sparsity is often too
restricting or unnatural, and weakened sparsity notions need to be
taken into account. On the other hand, sometimes -- such as with the
tree structure of wavelet coefficients -- some structural information
on the non-zero coefficients is known, which leads to diverse
structured sparsity models. Section \ref{sec:signalmodels} provides
an overview of such models.

\subsection{Recovery Algorithms: Optimization Theory and More}
\label{subsec:optimization}

\begin{figure}[h]
\centering
\includegraphics[width=6cm]{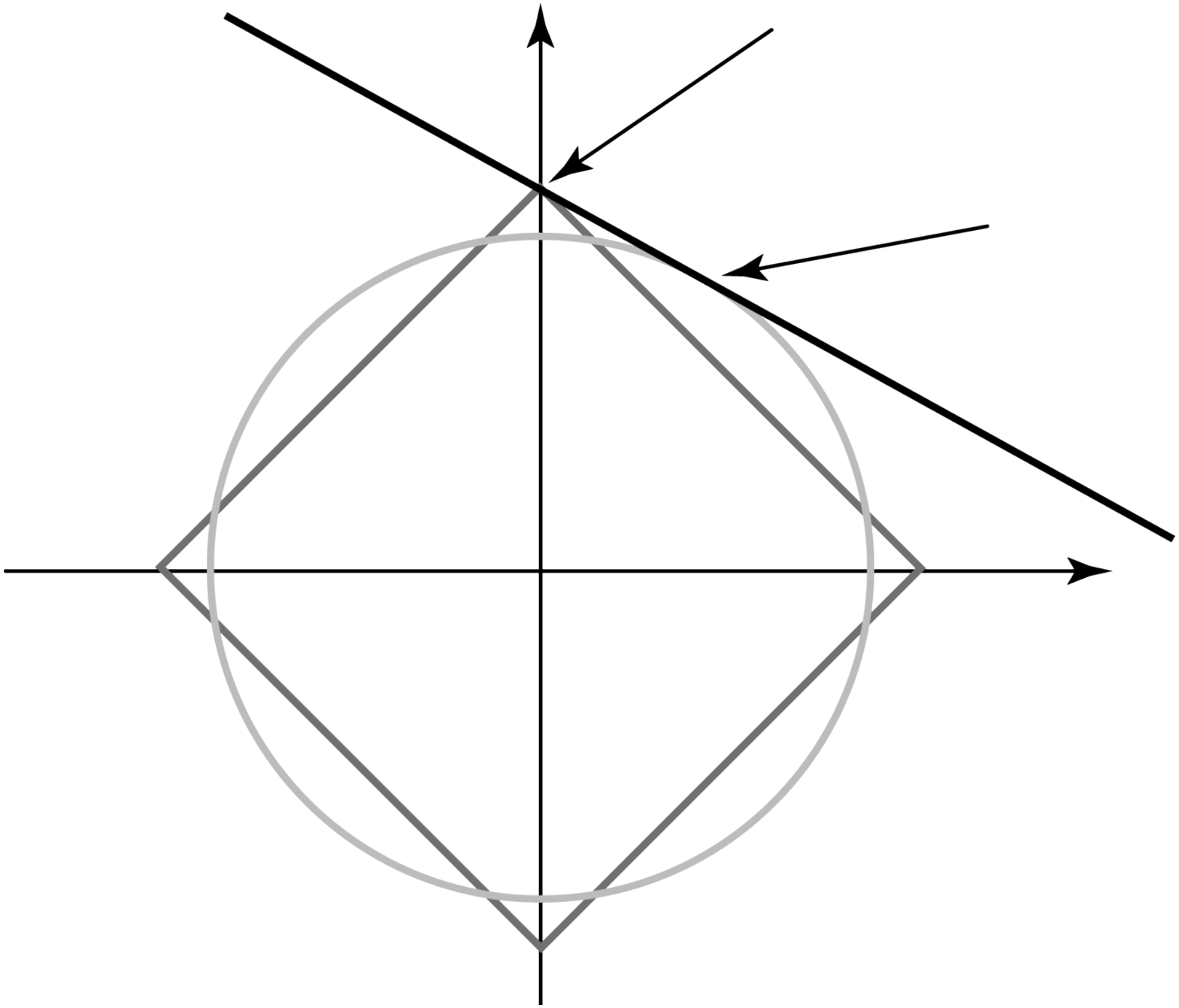}
\put(-12,53){\footnotesize $\{x : y = Ax\}$}
\put(-35,93){\footnotesize $\min \|x\|_2 \mbox{ s.t. } y = Ax$}
\put(-60,117){\footnotesize $\min \|x\|_1 \mbox{ s.t. } y = Ax$}
\caption{$\ell_1$ minimization versus $\ell_2$ minimization}
\label{fig:l1versusl2}
\end{figure}

\begin{figure}[h]
\centering
\includegraphics[width=5cm]{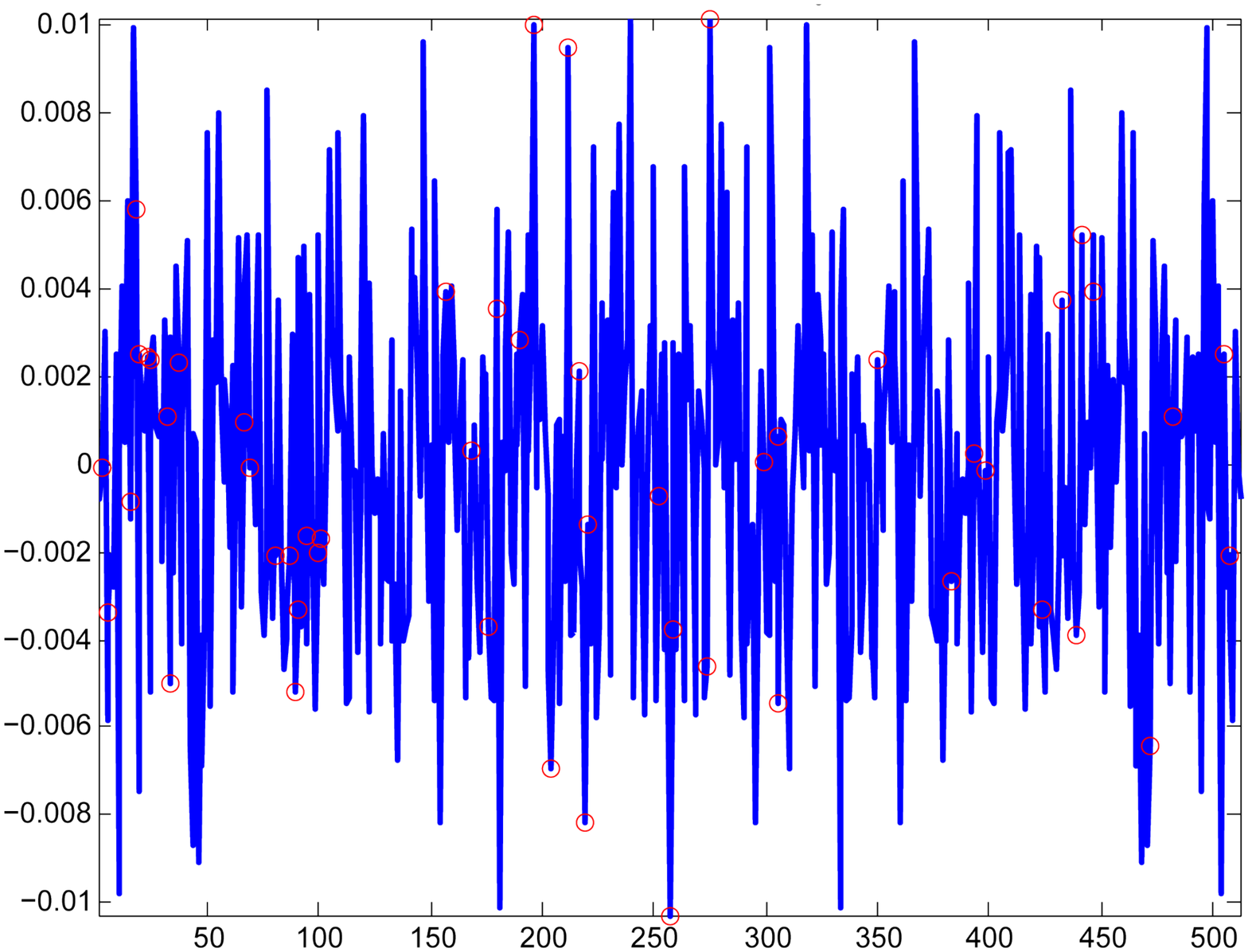}
\hspace*{0.2cm}
\includegraphics[width=5cm]{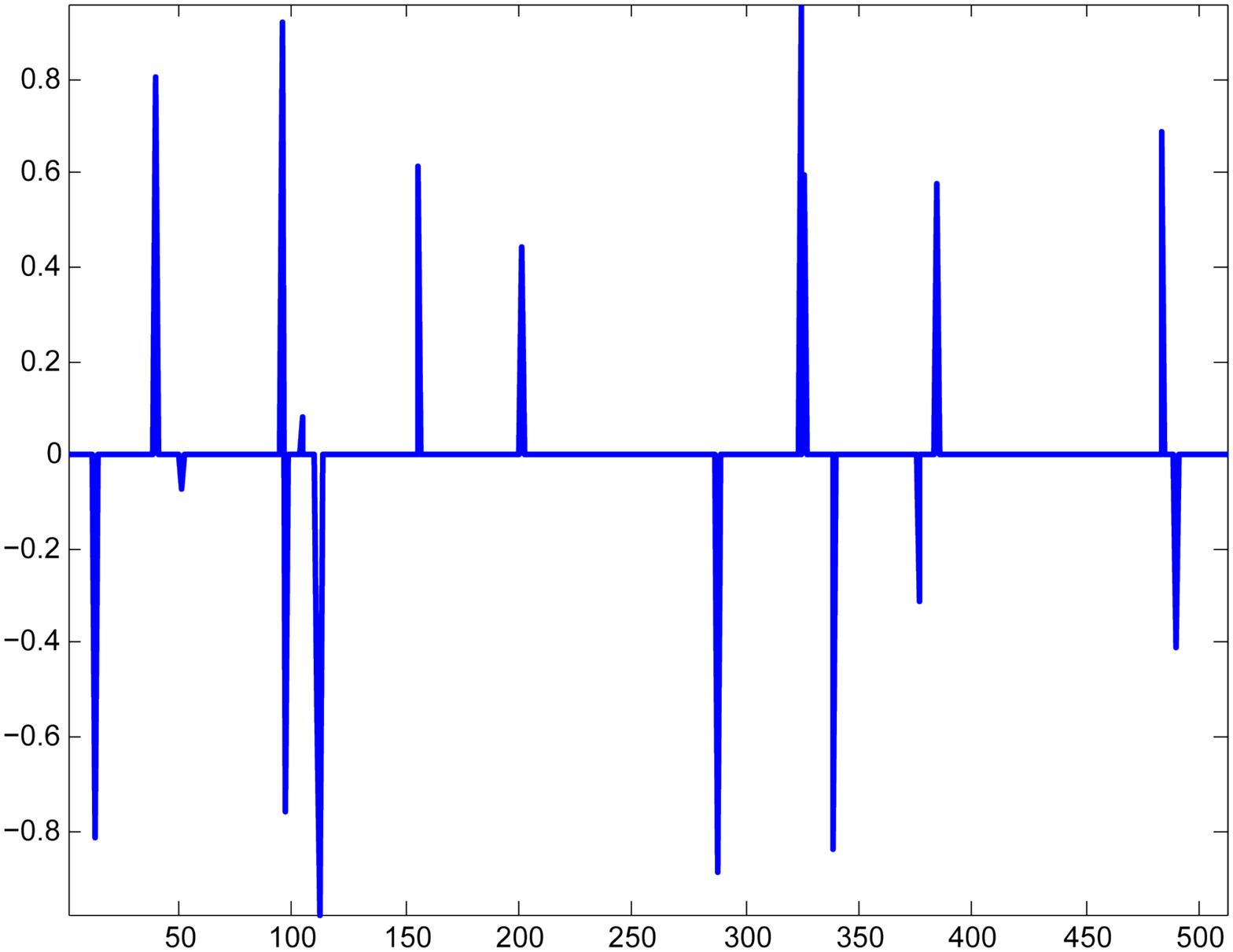}\\
\includegraphics[width=5cm]{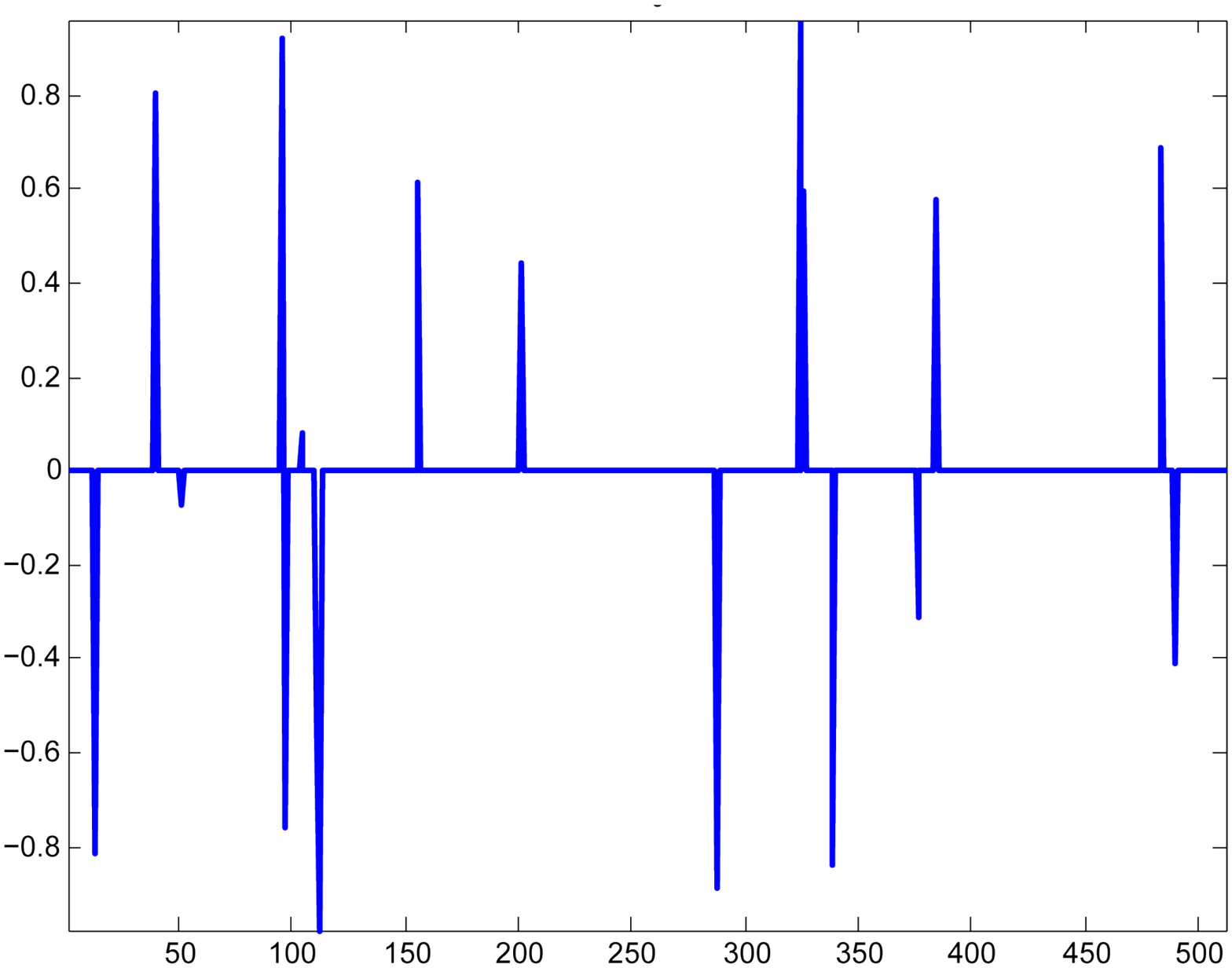}
\hspace*{0.2cm}
\includegraphics[width=5cm]{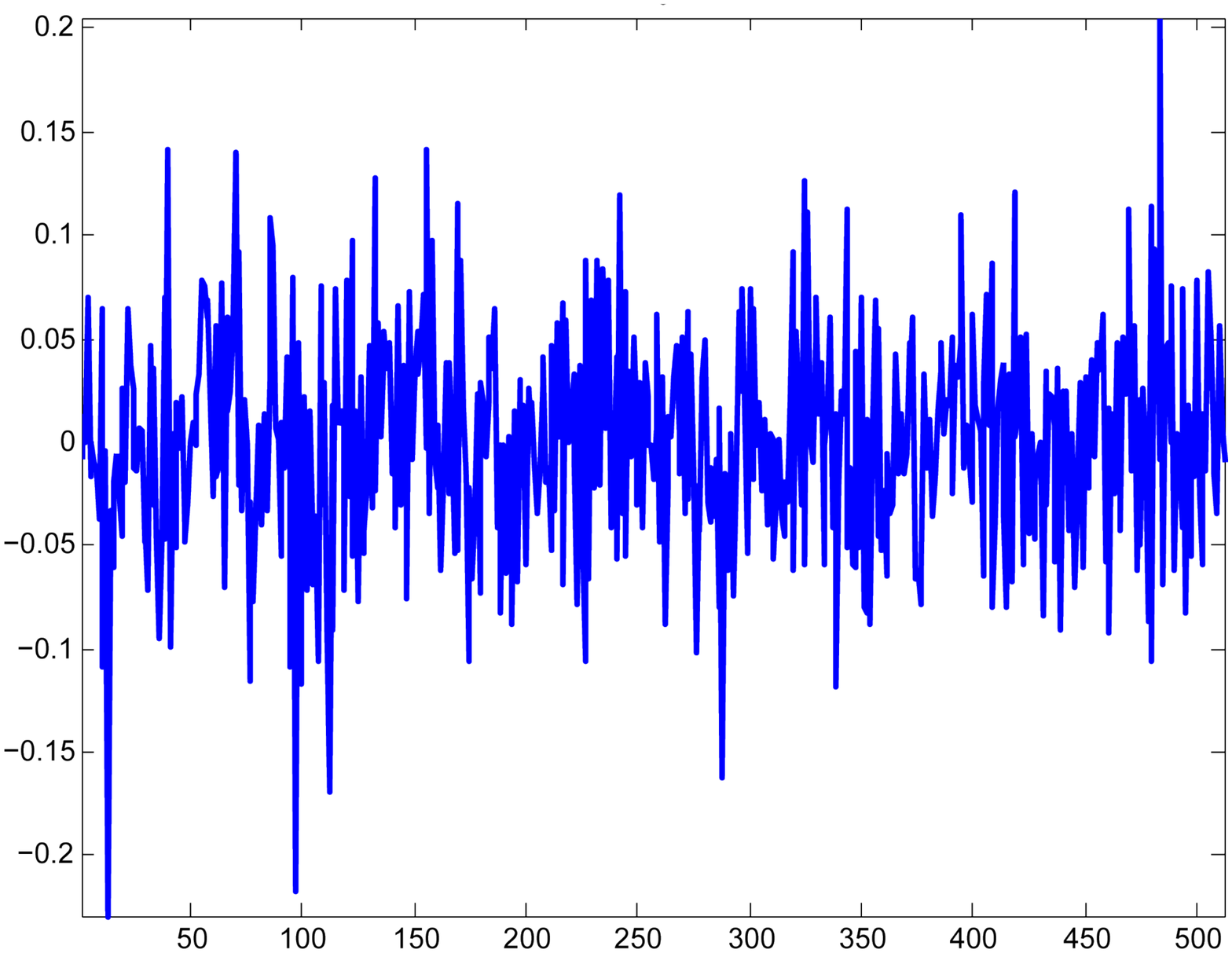}
\put(-227,113){(a)}
\put(-75,113){(b)}
\put(-227,-7){(c)}
\put(-75,-7){(d)}
\caption{(a) Original signal $f$ with random sample points (indicated by circles); (b) The Fourier transform $\hat{f}$;
(c) Perfect recovery of $\hat{f}$ by $\ell_1$ minimization; (d) Recovery of $\hat{f}$ by $\ell_2$ minimization}
\label{fig:l1versusl2_exp}
\end{figure}

Let $x$ now be a sparse vector. It is quite intuitive to recover $x$
from knowledge of $y$ by solving
\[
(P_0) \qquad \min_x \|x\|_0 \mbox{ subject to } y = Ax.
\]
Due to the unavoidable combinatorial search, this algorithm is
however NP-hard \cite{Mut05}. The main idea of Chen, Donoho, and
Saunders in the fundamental paper \cite{CDS98} was to substitute the
$\ell_0$ `norm' by the closest convex norm, which is the $\ell_1$
norm. This leads to the following minimization problem, which they
coined {\em Basis Pursuit}:
\[
(P_1) \qquad \min_x \|x\|_1 \mbox{ subject to } y = Ax.
\]
Due to the shape of the $\ell_1$ ball, $\ell_1$ minimization indeed
promotes sparsity. For an illustration of this fact, we refer the
reader to Figure \ref{fig:l1versusl2}, in which $\ell_1$ minimization
is compared to $\ell_2$ minimization. We would also like to draw the
reader's attention to the small numerical example in Figure \ref{fig:l1versusl2_exp},
in which a partial Fourier matrix is chosen as measurement matrix.

 The general question of when `$\ell_0 = \ell_1$'
holds is key to compressed sensing. Both necessary and sufficient
conditions have been provided, which not only depend on the sparsity
of the original vector $x$, but also on the incoherence of the
sensing matrix $A$, which will be made precise in
Section~\ref{sec:conditions}.

Since for very large data sets $\ell_1$ minimization is often not
feasible even when the solvers are adapted to the particular
structure of compressed sensing problems, various other types of
recovery algorithms were suggested. These can be roughly separated
into convex optimization, greedy, and combinatorial algorithms (cf.
Section \ref{sec:recoveryalgorithms}), each one having its own
advantages and disadvantages.

\subsection{Sensing Matrices: How Much Freedom is Allowed?}

As already mentioned, sensing matrices are required to satisfy
certain incoherence conditions such as, for instance, a small
so-called mutual coherence. If we are allowed to choose the sensing
matrix freely, the best choice are random matrices such as Gaussian
iid matrices, uniform random ortho-projectors, or Bernoulli
matrices, see for instance \cite{CRT06}.

It is still an open question (cf. Section \ref{sec:sensingmatrices}
for more details) whether deterministic matrices can be carefully
constructed to have similar properties with respect to compressed sensing problems. At the moment, different approaches towards this
problem are being taken such as structured random matrices by, for
instance, Rauhut et al.~in \cite{PRT11} or \cite{RRT12}. Moreover,
most applications do not allow for a free choice of the sensing matrix
and enforce a particularly structured matrix. Exemplary situations
are the application of data separation, in which the sensing matrix
has to consist of two or more orthonormal bases or frames
\cite[Chapter 11]{EK12}, or high resolution radar, for which the
sensing matrix has to bear a particular time-frequency structure
\cite{HS09}.

\subsection{Compressed Sensing: Quo Vadis?}
\label{subsec:quovadis}

At present, a comprehensive core theory seems established except for
some few deep questions such as the construction of deterministic
sensing matrices exhibiting properties similar to random matrices.

One current main direction of research which can be identified with
already various existing results is the incorporation of additional
sparsity properties typically coined {\em structured sparsity}, see
Section \ref{sec:signalmodels} for references. Another main
direction is the extension or transfer of the Compressed Sensing
Problem to other settings such as {\em matrix completion}, see for
instance \cite{CR08}. Moreover, we are currently witnessing the diffusion
of compressed sensing ideas to various {\em application areas} such
as radar analysis, medical imaging, distributed signal processing,
and data quantization, to name a few; see \cite{EK12} for an overview.
These applications pose intriguing challenges to the area due to the
constraints they require, which in turn initiates novel theoretical
problems. Finally, we observe that due to the need of, in
particular, fast sparse recovery algorithms, there is a trend to
more closely cooperate with {\em mathematicians from other research
areas}, for example from optimization theory, numerical linear
algebra, or random matrix theory.

As three examples of recently initiated research directions, we
would like to mention the following. First, while the theory of
compressed sensing focusses on digital data, it is desirable to
develop a similar theory for the {\em continuum setting}. Two
promising approaches were so far suggested by Eldar et al. (cf.
\cite{MEE11}) and Adcock et al. (cf. \cite{AH12}). Second, in
contrast to Basis Pursuit, which minimizes the $\ell_1$ norm of the
synthesis coefficients, several approaches such as recovery of
missing data minimize the $\ell_1$ norm of the analysis coefficients
-- as opposed to minimizing the $\ell_1$ norm of the synthesis
coefficients --, see Subsections \ref{subsec:frame1} and \ref{subsec:frame2}. The
relation between these two minimization problems is far from being
clear, and the recently introduced notion of {\em co-sparsity}
\cite{NDEG12} is an interesting approach to shed light onto this
problem. Third, the utilization of {\em frames as a sparsifying
system} in the context of compressed sensing has become a topic of
increased interest, and we refer to the initial paper \cite{CENP11}.

The reader might also want to consult the extensive webpage
\url{dsp.rice.edu/cs} containing most published papers in the area
of compressed sensing subdivided into different topics. We would
also like to draw the reader's attention to the recent books
\cite{Ela10} and \cite{EK12} as well as the survey article
\cite{BDE09}.

\subsection{Outline}

In Section \ref{sec:signalmodels}, we start by discussing different
sparsity models including structured sparsity and sparsifying
dictionaries. The next section, Section \ref{sec:conditions}, is
concerned with presenting both necessary and sufficient conditions
for exact recovery using $\ell_1$ minimization as a recovery strategy.
The delicateness of designing sensing matrices is the focus of
Section \ref{sec:sensingmatrices}. In Section
\ref{sec:recoveryalgorithms}, other algorithmic approaches to sparse
recovery are presented. Finally, applications such as data
separation are discussed in Section \ref{sec:applications}.


\section{Signal Models}
\label{sec:signalmodels}

Sparsity is the prior information assumed of the vector we intend to
efficiently sense or whose dimension we intend to reduce, depending
on which viewpoint we take. We will start by recalling some
classical notions of sparsity. Since applications typically impose a
certain structure on the significant coefficients, various
structured sparsity models were introduced which we will
subsequently present. Finally, we will discuss how to ensure
sparsity through an appropriate orthonormal basis or frame.

\subsection{Sparsity}

The most basic notion of sparsity states that a vector has at most $k$
non-zero coefficients. This is measured by the $\ell_0$ `norm',
which for simplicity we will throughout refer to as a norm although
it is well-known that $\|\cdot\|_0$ does not constitute a mathematical norm.

\begin{definition}\label{defi:sigma}
A vector $x = (x_i)_{i=1}^n \in \RR^n$ is called {\em $k$-sparse},
if
\[
\|x\|_0 = \#\{i : x_i \neq 0\} \le k.
\]
The set of all $k$-sparse vectors is denoted by $\Sigma_k$.
\end{definition}

We wish to emphasize that $\Sigma_k$ is a highly non-linear set. Letting $x \in \RR^n$
be a $k$-sparse signal, it belongs to the linear subspace consisting of all vectors with
the same support set. Hence the set $\Sigma_k$ is the union of all subspaces of vectors
with support $\Lambda$ satisfying $|\Lambda| \le k$.

From an application point of view, the situation of $k$-sparse vectors is however unrealistic,
wherefore various weaker versions were suggested. In the following
definition we present one possibility but do by no means claim this
to be the most appropriate one. It might though be very natural,
since it analyzes the decay rate of the $\ell_p$ error of the best
$k$-term approximation of a vector.

\begin{definition}
\label{defi:compressible} Let $1 \le p < \infty$ and $r > 0$. A vector $x = (x_i)_{i=1}^n \in \RR^n$ is
called {\em $p$-compressible with constant $C$ and rate $r$}, if
\[
\sigma_k(x)_p:= \min_{\tilde{x} \in \Sigma_k} \|x-\tilde{x}\|_p \le
C \cdot k^{-r} \quad \mbox{for any }k \in \{1, \ldots, n\}.
\]
\end{definition}

\subsection{Structured Sparsity}

Typically, the non-zero or significant coefficients do not arise in
arbitrary patterns but are rather highly structured. Think of the
coefficients of a wavelet decomposition which exhibit a tree
structure, see also Figure \ref{fig:sparsity}. To take these
considerations into account, structured sparsity models were
introduced. A first idea might be to identify the clustered set of
significant coefficients \cite{DK10}. An application of this notion
will be discussed in Section \ref{sec:applications}.

In the following definition as well as in the sequel, for some vector $x = (x_i)_{i=1}^n \in \RR^n$
and some subset $\Lambda \subset \{1, \ldots, n\}$, the expression $1_\Lambda x$ will
denote the vector in $\RR^n$ defined by
\[
(1_\Lambda x)_i = \left\{\begin{array}{rcl}x_i & : & i \in \Lambda,\\ 0 & : & i \not\in \Lambda, \end{array} \right.
\qquad i = 1, \ldots, n.
\]
Moreover, $\Lambda^c$ will denote the complement of the set $\Lambda$ in $\{1, \ldots, n\}$.

\begin{definition}
\label{defi:relativesparse} Let $\Lambda \subset \{1, \ldots, n\}$
and $\delta > 0$. A vector $x = (x_i)_{i=1}^n \in \RR^n$ is then
called {\em $\delta$-relatively sparse with respect to $\Lambda$},
if
\[
\|1_{\Lambda^c} x\|_1 \le \delta.
\]
\end{definition}

The notion of $k$-sparsity can also be regarded from a more general viewpoint, which simultaneously
imposes additional structure. Let $x \in \RR^n$ be a $k$-sparse signal. Then it belongs to the union
of linear one-dimensional subspaces consisting of all vectors with exactly one non-zero entry; this
entry belonging to the support set of $x$. The number of such subspaces equals $k$. Thus, a natural
extension of this concept is the following definition, initially
introduced in \cite{LD08}.

\begin{definition}
Let $(\cW_j)_{j=1}^N$ be a family of subspaces in $\RR^n$. Then a
vector $x \in \RR^n$ is {\em $k$-sparse in the union of subspaces
$\bigcup_{j=1}^N \cW_j$}, if there exists $\Lambda \subset \{1,
\ldots, N\}$, $|\Lambda| \le k$, such that
\[
x \in \bigcup_{j \in \Lambda} \cW_j.
\]
\end{definition}

At about the same time, the notion of {\em fusion frame sparsity}
was introduced in \cite{BKR11}. Fusion frames are a set of subspaces
having frame-like properties, thereby allowing for stability
considerations. A family of subspaces $(\cW_j)_{j=1}^N$ in $\RR^n$
is a {\em fusion frame} with bounds $A$ and $B$, if
\[
A \|x\|_2^2 \le \sum_{j=1}^N \|P_{\cW_j}(x)\|_2^2 \le B \|x\|_2^2
\quad \mbox{for all } x \in \RR^n,
\]
where $P_{\cW_j}$ denotes the orthogonal projection onto the
subspace $\cW_j$, see also \cite{CKL08} and \cite[Chapter 13]{CK12}.
Fusion frame theory extends classical frame theory by allowing the
analysis of signals through projections onto arbitrary dimensional
subspaces as opposed to one-dimensional subspaces in frame theory,
hence serving also as a model for distributed processing, cf. \cite{RJ05}. Fusion
frame sparsity can then be defined in a similar way as for a union
of subspaces.

Applications such as manifold learning assume that the signal under
consideration lives on a general manifold, thereby forcing us to
leave the world of linear subspaces. In such cases, the signal class
is often modeled as a non-linear $k$-dimensional manifold $\cM$ in
$\RR^n$, i.e.,
\[
x \in \cM = \{f(\theta) : \theta \in \Theta\}
\]
with $\Theta$ being a $k$-dimensional parameter space. Such signals
are then considered {\em $k$-sparse in the manifold model}, see
\cite{XH12}. For a survey chapter about this topic, the interested
reader is referred to \cite[Chapter 7]{EK12}.

\medskip

We wish  to finally mention that applications such as matrix
completion require generalizations of vector sparsity by
considering, for instance, low-rank matrix models. This is however
beyond the scope of this survey paper, and we refer to \cite{EK12}
for more details.

\subsection{Sparsifying Dictionaries and Dictionary Learning}

If the vector itself does not exhibit sparsity, we are required to
sparsify it by choosing an appropriate representation system -- in
this field typically coined {\em dictionary}. This problem can be
attacked in two ways, either non-adaptively or adaptively.

If certain characteristics of the signal are known, a dictionary can
be chosen from the vast class of already very well explored
representation systems such as the Fourier basis, wavelets, or
shearlets, to name a few. The achieved sparsity might not be
optimal, but various mathematical properties of these systems are
known and fast associated transforms are available.

Improved sparsity can be achieved by choosing the dictionary
adaptive to the signals at hand. For this, a test set of signals is
required, based on which a dictionary is learnt. This process is
customarily termed {\em dictionary learning}. The most well-known
and widely used algorithm is the K-SVD algorithm introduced by
Aharon, Elad, and Bruckstein in \cite{AEB06}. However, from a
mathematician's point of view, this approach bears two problems
which will hopefully be both solved in the near future.
First, almost no convergence results for such algorithms are
known. And, second, the learnt dictionaries do not exhibit any
mathematically exploitable structure, which makes not only an
analysis very hard but also prevents the design of fast associated
transforms.


\section{Conditions for Sparse Recovery}
\label{sec:conditions}

After having introduced various sparsity notions, in this sense
signal models, we next consider which conditions we need to impose
on the sparsity of the original vector and on the sensing matrix for
exact recovery. For the sparse recovery method, we will focus on $\ell_1$
minimization similar to most published results and refer to Section
\ref{sec:recoveryalgorithms} for further algorithmic approaches. In
the sequel of the present section, several incoherence conditions
for sensing matrices will be introduced. Section
\ref{sec:sensingmatrices} then discusses examples of matrices
fulfilling those. Finally, we mention that most results can be
slightly modified to incorporate measurements affected by additive
noise, i.e., if $y = Ax + \nu$ with $\|\nu\|_2 \le \varepsilon$.

\subsection{Uniqueness Conditions for Minimization Problems}

We start by presenting conditions for uniqueness of the solutions to the minimization
problems ($P_0$) and ($P_1$) which we introduced in Subsection
\ref{subsec:optimization}.

\subsubsection{Uniqueness of ($P_0$)}

The correct condition on the sensing matrix is phrased in terms of
the so-called spark, whose definition we first recall. This notion
was introduced in \cite{DE03} and verbally fuses the notions of
`sparse' and `rank'.

\begin{definition}
Let $A$ be an $m \times n$ matrix. Then the {\em spark} of $A$
denoted by spark($A$) is the minimal number of linearly dependent
columns of $A$.
\end{definition}

It is useful to reformulate this notion in terms of the null space
of $A$, which we will throughout denote by $\cN(A)$, and state its
range. The proof is obvious. For the definition of $\Sigma_k$, we
refer to Definition \ref{defi:sigma}.

\begin{lemma}
\label{lemm:spark} Let $A$ be an $m \times n$ matrix. Then
\[
\mbox{\rm spark}(A) = \min \{k : \cN(A) \cap \Sigma_k \neq \{0\}\}
\]
and {\rm spark($A$)} $\in [2,m+1]$.
\end{lemma}

This notion enables us to derive an equivalent condition on unique
solvability of ($P_0$). Since the proof is short, we state it for
clarity purposes.

\begin{theorem}[\cite{DE03}]
Let $A$ be an $m \times n$ matrix, and let $k \in \NN$. Then the
following conditions are equivalent. \bitem
\item[(i)] If a solution $x$ of ($P_0$) satisfies $\|x\|_0 \le k$, then this is the unique solution.
\item[(ii)] $k < \mbox{\rm spark($A$)}/2$.
\eitem
\end{theorem}

\noindent {\bf Proof.}
(i) $\Rightarrow$ (ii). We argue by
contradiction. If (ii) does not hold, by Lemma \ref{lemm:spark},
there exists some $h \in \cN(A)$, $h \neq 0$ such that $\|h\|_0 \le
2k$. Thus, there exist $x$ and $\tilde{x}$ satisfying $h =
x-\tilde{x}$ and $\|x\|_0, \|\tilde{x}\|_0 \le k$, but
$Ax=A\tilde{x}$, a contradiction to (i).

(ii) $\Rightarrow$ (i). Let $x$ and $\tilde{x}$ satisfy $y =
Ax=A\tilde{x}$ and $\|x\|_0, \|\tilde{x}\|_0 \le k$. Thus
$x-\tilde{x} \in \cN(A)$ and $\|x-\tilde{x}\|_0 \le 2k < \mbox{\rm
spark($A$)}$. By Lemma \ref{lemm:spark}, it follows that
$x-\tilde{x} = 0$, which implies (i).
\qed

\subsubsection{Uniqueness of $(P_1)$}

Due to the underdeterminedness of $A$ and hence the ill-posedness of
the recovery problem, in the analysis of uniqueness of the
minimization problem  $(P_1)$, the null space of $A$ also plays a
particular role. The related so-called null space property, first
introduced in \cite{CDD09}, is defined as follows.

\begin{definition}
Let $A$ be an $m \times n$ matrix. Then $A$ has the {\em null space
property (NSP) of order $k$}, if, for all $h \in \cN(A) \setminus
\{0\}$ and for all index sets $|\Lambda| \le k$,
\[
\|1_\Lambda h\|_1 < \tfrac12 \|h\|_1.
\]
\end{definition}

An equivalent condition for the existence of a unique sparse
solution of ($P_1$) can now be stated in terms of the null space
property. For the proof, we refer to \cite{CDD09}.

\begin{theorem}[\cite{CDD09}]
Let $A$ be an $m \times n$ matrix, and let $k \in \NN$. Then the
following conditions are equivalent. \bitem
\item[(i)] If a solution $x$ of ($P_1$) satisfies $\|x\|_0 \le k$, then this is the unique solution.
\item[(ii)] $A$ satisfies the null space property of order $k$.
\eitem
\end{theorem}

It should be emphasized that \cite{CDD09} studies the Compressed
Sensing Problem in a much more general way by analyzing quite
general encoding-decoding strategies.

\subsection{Sufficient Conditions}

The core of compressed sensing is to determine when `$\ell_0 =
\ell_1$', i.e., when the solutions of ($P_0$) and ($P_1$) coincide.
The most well-known sufficient conditions for this to hold true are
phrased in terms of mutual coherence and of the restricted isometry
property.

\subsubsection{Mutual Coherence}
\label{subsubsec:coherence}

The mutual coherence of a matrix, initially introduced in
\cite{DH01}, measures the smallest angle between each pair of its
columns.

\begin{definition}
Let $A = (a_i)_{i=1}^n$ be an $m \times n$ matrix. Then its {\em
mutual coherence} $\mu(A)$ is defined as
\[
\mu(A) = \max_{i \neq j} \frac{|\ip{a_i}{a_j}|}{\|a_i\|_2
\|a_j\|_2}.
\]
\end{definition}

The maximal mutual coherence of a matrix certainly equals $1$ in the case when
two columns are linearly dependent. The lower bound presented in the next result,
also known as the {\em Welch bound}, is more interesting. It can be
shown that it is attained by so-called {\em optimal Grassmannian frames}
\cite{SH04}, see also Section \ref{sec:sensingmatrices}.

\begin{lemma}
Let $A$ be an $m \times n$ matrix. Then we have
\[
\mu(A) \in \Big[\sqrt{\frac{n-m}{m(n-1)}},1\Big].
\]
\end{lemma}

Let us mention that different variants of mutual coherence exist, in
particular, the {\em Babel function} \cite{DE03}, the {\em
cumulative coherence function} \cite{Tro04}, the {\em structured
$p$-Babel function} \cite{BGN08}, the {\em fusion coherence}
\cite{BKR11}, and {\em cluster coherence} \cite{DK10}. The notion of
cluster coherence will in fact be later discussed in Section
\ref{sec:applications} for a particular application.

Imposing a bound on the sparsity of the original vector by the
mutual coherence of the sensing matrix, the following result can be
shown; its proof can be found in \cite{DE03}.

\begin{theorem}[\cite{EB02,DE03}]
\label{theo:coherence} Let $A$ be an $m \times n$ matrix, and let $x
\in \RR^n \setminus \{0\}$ be a solution of ($P_0$) satisfying
\[
\|x\|_0 < \tfrac{1}{2} (1+\mu(A)^{-1}).
\]
Then $x$ is the unique solution of ($P_0$) and ($P_1$).
\end{theorem}

\subsubsection{Restricted Isometry Property}

We next discuss the restricted isometry property, initially
introduced in \cite{CRT06}. It measures the degree to which each
submatrix consisting of $k$ column vectors of $A$ is close to being an isometry.
Notice that this notion automatically ensures stability, as will
become evident in the next theorem.

\begin{definition}
Let $A$ be an $m \times n$ matrix. Then $A$ has the {\em Restricted
Isometry Property (RIP) of order $k$}, if there exists a $\delta_k
\in (0,1)$ such that
\[
(1-\delta_k)\|x\|_2^2 \le \|Ax\|_2^2 \le  (1+\delta_k)\|x\|_2^2
\quad \mbox{for all } x \in \Sigma_k.
\]
\end{definition}

Several variations of this notion were introduced during the last
years, of which examples are the {\em fusion RIP} \cite{BKR11} and
the {\em D-RIP} \cite{CENP11}.

Although also for mutual coherence, error estimates for recovery
from noisy data are known, in the setting of the RIP those are very
natural. In fact, the error can be phrased in terms of the best
$k$-term approximation (cf. Definition \ref{defi:compressible}) as
follows.

\begin{theorem}[\cite{Can08,CDD09}]
Let $A$ be an $m \times n$ matrix which satisfies the RIP of order
$2k$ with $\delta_{2k} < \sqrt{2}-1$. Let $x \in \RR^n$, and let
$\hat{x}$ be a solution of the associated ($P_1$) problem. Then
\[
\|x-\hat{x}\|_2 \le C \cdot
\Big(\frac{\sigma_k(x)_1}{\sqrt{k}}\Big)
\]
for some constant $C$ dependent on $\delta_{2k}$.
\end{theorem}

The best known RIP condition for sparse recovery by ($P_1$) states
that ($P_1$) recovers all $k$-sparse vectors provided the measurement
matrix $A$ satisfies $\delta_{2k} < 0.473$, see \cite{Fou10}.

\subsection{Necessary Conditions}

Meaningful necessary conditions for `$\ell_0 = \ell_1$' in the sense of ($P_0$) $=$ ($P_1$) are significantly harder to
achieve. An interesting string of research was initiated by Donoho
and Tanner with the two papers \cite{DT05a,DT05b}. The main idea is
to derive equivalent conditions utilizing the theory of convex
polytopes. For this, let $C^n$ be defined by
\beq \label{eq:Cn}
C^n = \{x \in \RR^n : \|x\|_1 \le 1\}.
\eeq
A condition equivalent to `$\ell_0 = \ell_1$' can then be formulated
in terms of properties of a particular related polytope. For the relevant notions
from polytope theory, we refer to \cite{Gru03}.

\begin{theorem}[\cite{DT05a,DT05b}]
Let $C^n$ be defined as in \eqref{eq:Cn}, let $A$ be an $m
\times n$ matrix, and let the polytope $P$ be defined by $P = AC^n
\subseteq \RR^m$. Then the following conditions are equivalent.
\bitem
\item[(i)] The number of $k$-faces of $P$ equals the number of $k$-faces of $C^n$.
\item[(ii)] ($P_0$) $=$ ($P_1$).
\eitem
\end{theorem}


The geometric intuition behind this result is the fact that the
number of $k$-faces of $P$ equals the number of indexing sets
$\Lambda \subseteq \{1, \ldots, n\}$ with $|\Lambda| = k$ such that
vectors $x$ satisfying $\supp x = \Lambda$ can be recovered via
($P_1$).

Extending these techniques, Donoho and Tanner were also able to
provide highly accurate analytical descriptions of the occurring
phase transition when considering the area of exact recovery
dependent on the ratio of the number of equations to the number of
unknowns $n/m$ versus the ratio of the number of nonzeros to the
number of equations $k/n$. The interested reader is referred to \cite{DT09} for further details.


\section{Sensing Matrices}
\label{sec:sensingmatrices}

Ideally, we aim for a matrix which has high spark, low mutual coherence,
and a small RIP constant. As our discussion in this section will show, these
properties are often quite difficult to achieve, and even computing, for
instance, the RIP constant is computationally intractable in general (see \cite{PT12}).

In the sequel, after presenting some general relations between the introduced
notions of spark, NSP, mutual coherence, and RIP, we will discuss
some explicit constructions for, in particular, mutual coherence and
RIP.

\subsection{Relations between Spark, NSP, Mutual Coherence, and RIP}

Before discussing different approaches to construct a sensing
matrix, we first present several known relations between the
introduced notions spark, NSP, mutual coherence, and RIP. This
allows to easily compute or at least estimate other measures, if a
sensing matrix is designed for a particular measure. For the proofs of
the different statements, we refer to \cite[Chapter 1]{EK12}.

\begin{lemma}
Let $A$ be an $m \times n$ matrix with normalized columns. \bitem
\item[(i)] We have
\[
\mbox{\rm spark}(A) \ge 1 + \frac{1}{\mu(A)}.
\]
\item[(ii)] $A$ satisfies the RIP of order $k$ with $\delta_k=k\mu(A)$ for all
$k < \mu(A)^{-1}$.
\item[(iii)] Suppose $A$ satisfies the RIP of order $2k$ with $\delta_{2k} < \sqrt{2}-1$. If
\[
\frac{\sqrt{2}\delta_{2k}}{1-(1+\sqrt{2})\delta_{2k}}  <
\sqrt{\frac{k}{n}},
\]
then $A$ satisfies the NSP of order $2k$. \eitem
\end{lemma}

\subsection{Spark and Mutual Coherence}

Let us now provide some exemplary classes of sensing matrices with
advantageous spark and mutual coherence properties.

The first observation one can make (see also \cite{CDD09}) is that
an $m \times n$ Vandermonde matrix $A$ satisfies
\[
\mbox{\rm spark}(A) = m+1.
\]
One serious drawback though is the fact that these matrices become
badly conditioned as $n \to \infty$.

Turning to the weaker notion of mutual coherence, of particular
interest -- compare Subsection \ref{subsec:separation} -- are
sensing matrices composed of two orthonormal bases or frames for
$\RR^m$. If the two orthonormal bases $\Phi_1$ and $\Phi_2$, say,
are chosen to be mutually unbiased such as the Fourier and the Dirac
basis (the standard basis), then
\[
\mu([\Phi_1 | \Phi_2]) = \frac{1}{\sqrt{m}},
\]
which is the optimal bound on mutual coherence for such types of $m
\times 2m$ sensing matrix. Other constructions are known for $m
\times m^2$ matrices $A$ generated from the Alltop sequence
\cite{HS09} or by using Grassmannian frames \cite{SH04}, in which
cases the optimal lower bound is attained:
\[
\mu(A) = \frac{1}{\sqrt{m}}.
\]
The number of measurements required for recovery of a $k$-sparse
signal can then be determined to be $m = O(k^2 \log n)$.

\subsection{RIP}

We begin by discussing some deterministic constructions of matrices
satisfying the RIP. The first noteworthy construction was presented
by DeVore and requires $m \gtrsim k^2$, see \cite{Dev07}. A very
recent, highly sophisticated approach  \cite{BDFKK11} by Bourgain et
al. still requires $m \gtrsim k^{2-\alpha}$ with some small constant
$\alpha$. Hence up to now deterministic constructions require a
large $m$, which is typically not feasible for applications, since
it scales quadratically in $k$.

The construction of random sensing matrices satisfying RIP is a
possibility to circumvent this problem. Such constructions are
closely linked to the famous Johnson-Lindenstrauss Lemma, which is
extensively utilized in numerical linear algebra, machine learning,
and other areas requiring dimension reduction.

\begin{theorem}[Johnson-Lindenstrauss Lemma \cite{JL84}]
Let $\varepsilon \in (0,1)$, let $x_1, \ldots, x_p \in \RR^n$, and
let $m = O(\varepsilon^{-2} \log p)$ be a positive integer. Then
there exists a Lipschitz map $f : \RR^n \to \RR^m$ such that
\[
(1-\varepsilon)\|x_i-x_j\|_2^2 \le \|f(x_i)-f(x_j)\|_2^2 \le
(1+\varepsilon)\|x_i-x_j\|_2^2 \quad \mbox{for all } i,j \in \{1,
\ldots, p\}.
\]
\end{theorem}

The key requirement for a matrix to satisfy the Johnson-Lindenstrauss Lemma with high probability
is the following concentration inequality for an arbitrarily fixed $x \in \RR^n$:
\beq \label{eq:5}
\mathbb{P}\Big((1-\varepsilon)\|x\|_2^2 \le \|Ax\|_2^2 \le (1+\varepsilon)\|x\|_2^2\Big) \le 1-2e^{-c_0\varepsilon^2m},
 \eeq
with the entries of $A$ being generated by a certain probability distribution.
The relation of RIP to the  Johnson-Lindenstrauss Lemma is
established in the following result. We also mention that recently
even a converse of the following theorem was proved in \cite{KW11}.

\begin{theorem}[\cite{BDDW08}]
Let $\delta \in (0,1)$.  If the probability distribution generating
the $m \times n$ matrices $A$ satisfies the concentration inequality
\eqref{eq:5} with $\varepsilon = \delta$, then there exist constants
$c_1, c_2$ such that, with probability $\le 1-2e^{-c_2\delta^2m}$,
$A$ satisfies the RIP of order $k$ with $\delta$ for all $k \le c_1
\delta^2 m/\log(n/k)$.
\end{theorem}

This observation was then used in \cite{BDDW08} to prove that
Gaussian and Bernoulli random matrices satisfy the RIP of order $k$
with $\delta$ provided that $m \gtrsim \delta^{-2} k  \log(n/k)$.
Up to a constant, lower bounds for Gelfand widths of $\ell_1$-balls
\cite{FPRU10} show that this dependence on $k$ and $n$ is indeed
optimal.


\section{Recovery Algorithms}
\label{sec:recoveryalgorithms}

In this section, we will provide a brief overview of the different
types of algorithms typically used for sparse recovery. Convex
optimization algorithms require very few measurements but are
computationally more complex. On the other extreme are combinatorial
algorithms, which are very fast -- often sublinear -- but require
many measurements that are sometimes difficult to obtain. Greedy
algorithms are in some sense a good compromise between those
extremes concerning computational complexity and the required number
of measurements.

\subsection{Convex Optimization}

In Subsection \ref{subsec:optimization}, we already stated the
convex optimization problem
\[
\min_x \|x\|_1 \;\mbox{ subject to }\;  y=Ax
\]
most commonly used. If the measurements are affected by noise, a
conic constraint is required; i.e., the minimization problem needs
to be changed to
\[
\min_x \|x\|_1 \;\mbox{ subject to }\;  \norm{Ax-y}_2^2 \le
\varepsilon,
\]
for a carefully chosen $\varepsilon > 0$. For a particular regularization parameter
$\lambda > 0$, this problem is equivalent to the unconstrained
version given by
\[
\min_x \tfrac12 \norm{Ax-y}_2^2 + \lambda \|x\|_1.
\]

Developed convex optimization algorithms specifically adapted to the
compressed sensing setting include interior-point methods
\cite{CRT06}, projected gradient methods \cite{FNW07}, and iterative
thresholding \cite{DDD04}. The reader might also be interested to
check the webpages \url{www-stat.stanford.edu/~candes/l1magic} and
\url{sparselab.stanford.edu} for available code.
It is worth pointing out that the intense research performed in this area has slightly diminished
the computational disadvantage of convex optimization algorithms for compressed sensing
as compared to greedy type algorithms.

\subsection{Greedy Algorithms}

Greedy algorithms iteratively approximate the coefficients and the
support of the original signal. They have the advantage of being
very fast and easy to implement. Often the theoretical performance
guarantees are very similar to, for instance, $\ell_1$ minimization
results.

\begin{figure}[h]
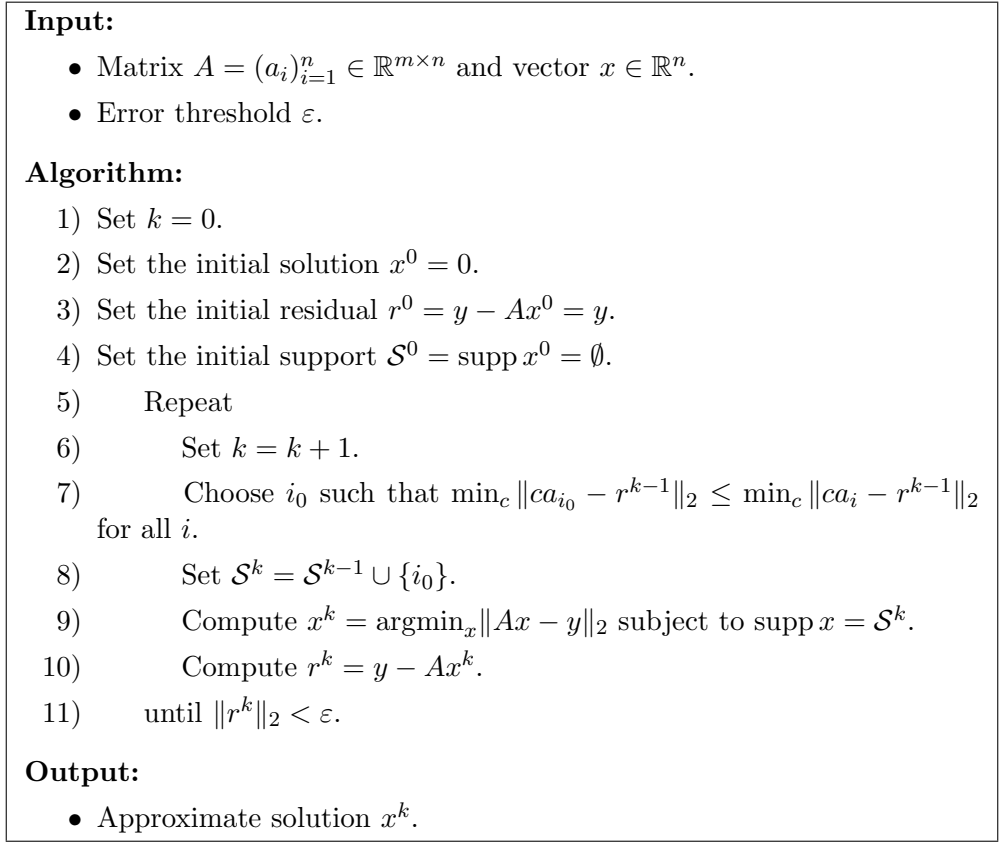

\centering \framebox{
\begin{minipage}[h]{5in}
{\bf Input:}\\[-4ex]
\begin{itemize}
\item Matrix $A = (a_i)_{i=1}^n \in \RR^{m \times n}$ and vector $x \in \RR^n$.\\[-4ex]
\item Error threshold $\varepsilon$.
\end{itemize}

{\bf Algorithm:}\\[-4ex]
\begin{itemize}
\item[1)] Set $k=0$.\\[-4ex]
\item[2)] Set the initial solution $x^0 = 0$.\\[-4ex]
\item[3)] Set the initial residual $r^0 = y-Ax^0 = y$.\\[-4ex]
\item[4)] Set the initial support $\cS^0 = \supp x^0 = \emptyset$.\\[-4ex]
\item[5)] \hspace*{0.5cm} Repeat\\[-4ex]
\item[6)] \hspace*{1cm} Set $k=k+1$.\\[-4ex]
\item[7)] \hspace*{1cm} Choose $i_0$ such that $\min_c \|c a_{i_0} - r^{k-1}\|_2 \le \min_c \|c a_{i} - r^{k-1}\|_2$
for all $i$.\\[-4ex]
\item[8)] \hspace*{1cm} Set $\cS^k = \cS^{k-1} \cup \{i_0\}$.\\[-4ex]
\item[9)] \hspace*{1cm} Compute $x^k = \argmin_x \|Ax-y\|_2$ subject to $\supp x = \cS^k$.\\[-4ex]
\item[10)] \hspace*{1cm} Compute $r^k = y - A x^k$.\\[-4ex]
\item[11)] \hspace*{0.5cm} until $\|r^k\|_2 < \varepsilon$.
\end{itemize}

{\bf Output:}\\[-4ex]
\begin{itemize}
\item Approximate solution $x^k$.
\end{itemize}
\end{minipage}
} \caption{Orthogonal Matching Pursuit (OMP): Approximation of the
solution of ($P_0$)} \label{fig:OMP}
\end{figure}

The most well-known greedy approach is {\em Orthogonal Matching
Pursuit}, which is described in Figure \ref{fig:OMP}. OMP was
introduced in \cite{PRK93} as an improved successor of {\em Matching
Pursuit} \cite{MZ93}.

Interestingly, a theorem similar to Theorem \ref{theo:coherence} can
be proven for OMP.

\begin{theorem}[\cite{Tro04,DET06}]
Let $A$ be an $m \times n$ matrix, and let $x \in \RR^n \setminus
\{0\}$ be a solution of ($P_0$) satisfying
\[
\|x\|_0 < \tfrac{1}{2} (1+\mu(A)^{-1}).
\]
Then OMP with error threshold $\varepsilon=0$ recovers $x$.
\end{theorem}

Other prominent examples of greedy algorithms are Stagewise OMP
(StOMP) \cite{DTDS07}, Regularized OMP (ROMP) \cite{NV09}, and
Compressive Sampling MP (CoSaMP) \cite{NT08}. For a survey of these
methods, we wish to refer to \cite[Chapter 8]{EK12}.

An intriguing, very recently developed class of algorithms is Orthogonal Matching
Pursuit with Replacement (OMPR) \cite{JTD12}, which not only includes most
iterative (hard)-thresholding algorithms as special cases, but
this approach also permits the tightest known analysis in terms
of RIP conditions. By extending OMPR using locality sensitive hashing (OMPR-Hash),
this also leads to the first provably sub-linear algorithm for sparse
recovery, see \cite{JTD12}. Another recent
development is message passing algorithms for compressed sensing
pioneered in \cite{DMM09}; a survey on those can be found in \cite[Chapter
9]{EK12}.

\subsection{Combinatorial Algorithms}

These methods apply group testing to highly structured samples of
the original signal, but are far less used in compressed sensing
as opposed to convex optimization and greedy algorithms. From the various types of algorithms, we
mention the HHS pursuit \cite{GSV07} and a sub-linear Fourier
transform \cite{Iwe10}.


\section{Applications}
\label{sec:applications}

We now turn to some applications of compressed sensing. Two of those
we will discuss in more detail, namely data separation and recovery
of missing data.

\subsection{Data Separation}
\label{subsec:separation}

The data separation problem can be stated in the following way. Let
$x=x_1+x_2 \in \RR^n$. Assuming we are just given $x$, how can we
extract $x_1$ and $x_2$ from it? At first glance, this seems to be
impossible, since there are two unknowns for one datum.

\subsubsection{An Orthonormal Basis Approach}

The first approach to apply compressed sensing techniques consists
in choosing appropriate orthonormal bases $\Phi_1$ and $\Phi_2$ for
$\RR^n$ such that the coefficient vectors $\Phi_i^T x_i$ ($i=1, 2$)
are sparse. This leads to the following underdetermined linear
system of equations:
\[
x = [\:\Phi_1\:|\:\Phi_2\:] \left[ \begin{array}{c} c_1\\ c_2
\end{array}\right].
\]
Compressed sensing now suggests to solve
\beq \label{eq:1}
\min_{c_1,c_2} \left\|\left[ \begin{array}{c} c_1\\ c_2
\end{array}\right]\right\|_1 \mbox{ subject to } x =
[\:\Phi_1\:|\:\Phi_2\:] \left[ \begin{array}{c} c_1\\ c_2
\end{array}\right].
\eeq
If the sparse vector $[\Phi_1^T x_1,\Phi_2^T x_2]^T$ can be recovered,
the data separation problem can be solved by computing
\[
x_1 = \Phi_1 (\Phi_1^T x_1) \quad \mbox{and} \quad x_2 = \Phi_2
(\Phi_2^T x_2).
\]
Obviously, separation can only be achieved provided that the
components $x_1$ and $x_2$ are in some sense morphologically
distinct. Notice that this property is indeed encoded in the problem
if one requires incoherence of the matrix
$[\:\Phi_1\:|\:\Phi_2\:]$.

In fact, this type of problem can be regarded as the birth of
compressed sensing, since the fundamental paper \cite{DH01} by
Donoho and Huo analyzed a particular data separation problem, namely
the separation of sinusoids and spikes. In this setting, $x_1$
consists of $n$ samples of a continuum domain signal which is a
superposition of sinusoids:
\[
x_1 = \left(\frac{1}{\sqrt{n}} \sum_{\omega = 0}^{n-1} c_{1,\omega}
e^{2 \pi i \omega t/n}\right)_{0 \le t \le n-1}
\]
Letting $\Phi_1$ be the Fourier basis, the coefficient vector
\[
\Phi_1^T x_1 = c_1, \quad \mbox{where } \Phi_1 = [\:\varphi_{1,0}\:|\:
\ldots \:|\: \varphi_{1,n-1}\:] \mbox{ with } \varphi_{1,\omega} =
\left(\tfrac{1}{\sqrt{n}} e^{2 \pi i \omega t/n}\right)_{0 \le t \le
n-1},
\]
is sparse. The vector $x_2$ consists of $n$ samples of a continuum
domain signal which is a superposition of spikes, i.e., has few
non-zero coefficients. Thus, letting $\Phi_2$ denote the Dirac
basis (standard basis), the coefficient vector
\[
\Phi_2^T x_2 = x_2 = c_2
\]
is also sparse. Since the mutual coherence of the matrix
$[\:\Phi_1\:|\:\Phi_2\:]$ can be computed to be
$\frac{1}{\sqrt{n}}$, Theorem \ref{theo:coherence} implies the
following result.

\begin{theorem}[\cite{DH01,EB02}]
Let $x_1, x_2$ and $\Phi_1, \Phi_2$ be defined as in the previous
paragraph, and assume that
$\|\Phi_1^T x_1\|_0 + \|\Phi_2^T x_2\|_0 < \tfrac{1}{2}
(1+\sqrt{n})$. Then
\[
\left[ \begin{array}{c} \Phi_1^T x_1\\ \Phi_2^T x_2
\end{array}\right] = \argmin_{c_1,c_2} \left\|\left[
\begin{array}{c} c_1\\ c_2 \end{array}\right]\right\|_1 \mbox{
subject to }x = [\:\Phi_1\:|\:\Phi_2\:] \left[ \begin{array}{c}
c_1\\ c_2 \end{array}\right].
\]
\end{theorem}

\subsubsection{A Frame Approach}
\label{subsec:frame1}

Now assume that we cannot find sparsifying orthonormal bases but
Parseval frames\footnote{Recall that $\Phi$ is a Parseval frame, if $\Phi \Phi^T = I$.} $\Phi_1$ and $\Phi_2$ -- notice that this situation
is much more likely due to the advantageous redundancy of a frame.
In this case, the minimization problem we stated in \eqref{eq:1}
faces the following problem: We are merely interested in the
separation $x=x_1+x_2$. However, for each such separation,  due to
the redundancy of the frames the minimization problem searches
through infinitely many coefficients $[c_1,c_2]^T$ satisfying
$x_i=\Phi_i c_i$, $i=1, 2$. Thus it computes not only much more than
necessary -- in fact, it even computes the sparsest coefficient
sequence of $x$ with respect to the dictionary
$[\:\Phi_1\:|\:\Phi_2\:]$ -- but this also causes numerical
instabilities if the redundancy of the frames is too high.

To avoid this problem, we place the $\ell_1$ norm on the {\em
analysis}, rather than the {\em synthesis} side as already mentioned
in Subsection \ref{subsec:quovadis}. Utilizing the fact that
$\Phi_1$ and $\Phi_2$ are Parseval frames, i.e., that $\Phi_i
\Phi_i^T = I$ ($i=1, 2$), we can write
\[
x=x_1+x_2=\Phi_1 (\Phi_1^T x_1) + \Phi_2 (\Phi_2^T x_2).
\]
This particular choice of coefficients -- which are in frame theory
language termed {\em analysis coefficients} -- leads to the
minimization problem \beq \label{eq:2}
\min_{\tilde{x}_1,\tilde{x}_2} \|\Phi_1^T \tilde{x}_1\|_1 +
\|\Phi_2^T \tilde{x}_2\|_1  \mbox{ subject to } x = \tilde{x}_1 +
\tilde{x}_2. \eeq Interestingly, the associated recovery results
employ structured sparsity, wherefore we will also briefly present
those. First, the notion of relative sparsity (cf. Definition
\ref{defi:relativesparse}) is adapted to this situation.

\begin{definition}
Let $\Phi_1$ and $\Phi_2$ be Parseval frames for $\RR^n$ with
indexing sets $\{1, \ldots, N_1\}$ and $\{1, \ldots, N_2\}$,
respectively, let $\Lambda_i \subset \{1, \ldots, N_i\}$, $i=1, 2$,
and let $\delta > 0$. Then the vectors $x_1$ and $x_2$ are called
{\em $\delta$-relatively sparse in $\Phi_1$ and $\Phi_2$ with
respect to $\Lambda_1$ and $\Lambda_2$}, if
\[
\|1_{\Lambda_1^c} \Phi_1^T x_1\|_1 + \|1_{\Lambda_2^c} \Phi_2^T
x_2\|_1 \le \delta.
\]
\end{definition}

Second, the notion of  mutual coherence is adapted to structured
sparsity as already discussed in Subsection \ref{subsubsec:coherence}.
This leads to the following definition of cluster coherence.

\begin{definition}
Let $\Phi_1=(\varphi_{1i})_{i=1}^{N_1}$ and
$\Phi_2=(\varphi_{2j})_{j=1}^{N_2}$ be Parseval frames for $\RR^n$,
respectively, and let $\Lambda_1 \subset \{1, \ldots, N_1\}$. Then
the {\em cluster coherence $\mu_c(\Lambda_1,\Phi_1;\Phi_2)$ of
$\Phi_1$ and $\Phi_2$ with respect to $\Lambda_1$} is defined by
\[
\mu_c(\Lambda_1,\Phi_1;\Phi_2) = \max_{j=1, \ldots, N_2} \sum_{i \in
\Lambda_1} \absip{\varphi_{1i}}{\varphi_{2j}}.
\]
\end{definition}

The performance of the minimization problem \eqref{eq:2} can then be
analyzed as follows. It should be emphasized that the clusters of
significant coefficients $\Lambda_1$ and $\Lambda_2$ are a mere
analysis tool; the algorithm does not take those into account.
Further, notice that the choice of those sets is highly delicate in
its impact on the separation estimate. For the proof of the result,
we refer to \cite{DK10}.

\begin{theorem} [\cite{DK10}]
Let $x=x_1+x_2 \in \RR^n$, let $\Phi_1$ and $\Phi_2$ be Parseval
frames for $\RR^n$ with indexing sets $\{1, \ldots, N_1\}$ and $\{1,
\ldots, N_2\}$, respectively, and let $\Lambda_i \subset \{1,
\ldots, N_i\}$, $i=1, 2$. Further, suppose that $x_1$ and $x_2$ are
$\delta$-relatively sparse in $\Phi_1$ and $\Phi_2$ with respect to
$\Lambda_1$ and $\Lambda_2$, and let $[x_1^\star,x_2^\star]^T$ be a
solution of the minimization problem \eqref{eq:2}. Then
\[
  \norm{x_1^\star-x_1}_2 + \norm{x_2^\star-x_2}_2 \le \frac{2\delta}{1-2\mu_c},
\]
where $\mu_c =
\max\{\mu_c(\Lambda_1,\Phi_1;\Phi_2),\mu_c(\Lambda_2,\Phi_2;\Phi_1)\}$.
\end{theorem}

\begin{figure}[h]
\centering
\includegraphics[width=7cm]{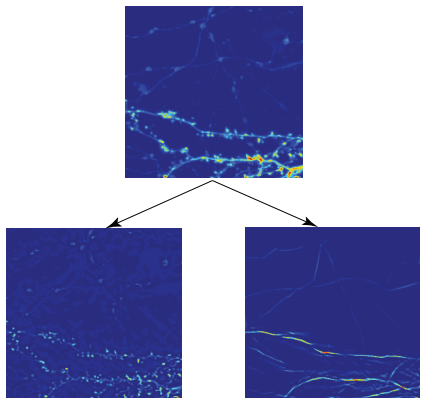}
\caption{Separation of a neurobiological image using wavelets and
shearlets \cite{KL10b}} \label{fig:separation}
\end{figure}

Let us finally mention that data separation via compressed sensing
has been applied, for instance, in imaging sciences for the
separation of point- and curvelike objects, a problem appearing in
several areas such as in astronomical imaging when separating stars
from filaments and in neurobiological imaging when separating spines
from dendrites. Figure \ref{fig:separation} illustrates a numerical
result from \cite{KL10b} using wavelets (see \cite{Mal98}) and
shearlets (see \cite{KL10a,KL12}) as sparsifying frames. A
theoretical foundation for separation of point- and curvelike objects
by $\ell_1$ minimization is developed in \cite{DK10}. When considering
thresholding as separation method for such features, even stronger
theoretical results could be proven in \cite{Kut12b}. Moreover, a first
analysis of separation of cartoon and texture -- very commonly
present in natural images -- was performed in \cite{Kut12a}.

For more details on data separation using compressed sensing
techniques, we refer to \cite[Chapter 11]{EK12}.

\subsection{Recovery of Missing Data}
\label{subsec:inpainting}

The problem of recovery of missing data can be formulated as
follows. Let $x = x_K + x_M \in \cW \oplus \cW^\perp$, where $\cW$
is a subspace of $\RR^n$. We assume only $x_K$ is known to us, and we
aim to recover $x$. Again, this seems unfeasible unless we have
additional information.

\subsubsection{An Orthonormal Basis Approach}

We now assume that -- although $x$ is not known to us -- we at least
know that it is sparsified by an orthonormal basis $\Phi$, say.
Letting $P_\cW$ and $P_{\cW^\perp}$ denote the orthogonal
projections onto $\cW$ and $\cW^\perp$, respectively, we are led to
solve the underdetermined problem
\[
P_\cW \Phi c = P_\cW x
\]
for the sparse solution $c$. As in the case of data separation, from
a compressed sensing viewpoint it is suggestive to solve \beq
\label{eq:3} \min_{c} \|c\|_1 \mbox{ subject to } P_\cW \Phi c =
P_\cW x. \eeq The original vector $x$ can then be recovered via $x =
\Phi c$. The solution of the inpainting problem -- a terminology used
for recovery of missing data in imaging science -- was first considered in
\cite{ESQD05}.

Application of Theorem \ref{theo:coherence} provides a sufficient
condition for missing data recovery to succeed.

\begin{theorem}[\cite{DE03}]
Let $x \in \RR^n$, let $\cW$ be a subspace of $\RR^n$, and let $\Phi$ be
an orthonormal basis for $\RR^n$. If
$\|\Phi^T x\|_0 < \frac12(1+\mu(P_\cW \Phi)^{-1})$, then
\[
\Phi^T x = \argmin_{c} \|c\|_1 \mbox{ subject to } P_\cW \Phi c =
P_\cW x.
\]
\end{theorem}

\subsubsection{A Frame Approach}
\label{subsec:frame2}

As before, we now assume that the sparsifying system $\Phi$ is a
redundant Parseval frame. The adapted version to \eqref{eq:3}, which
places the $\ell_1$ norm on the analysis side, reads \beq
\label{eq:4} \min_{\tilde{x}} \|\Phi^T \tilde{x}\|_1 \mbox{ subject
to } P_\cW \tilde{x} = P_\cW x. \eeq

Employing relative sparsity and cluster coherence, an error analysis
can be derived in a similar way as before. For the proof, the reader
might want to consult \cite{KKZ12}.

\begin{theorem}[\cite{KKZ12}]
Let $x \in \RR^n$, let $\Phi$ be a Parseval frame for $\RR^n$ with
indexing set $\{1, \ldots, N\}$, and let $\Lambda \subset \{1,
\ldots, N\}$. Further, suppose that $x$ is $\delta$-relatively
sparse in $\Phi$ with respect to $\Lambda$, and let $x^\star$ be a
solution of the minimization problem \eqref{eq:4}. Then
\[
  \norm{x^\star-x}_2 \le \frac{2\delta}{1-2\mu_c},
\]
where $\mu_c = \mu_c(\Lambda, P_{\cW^\perp} \Phi;\Phi)$.
\end{theorem}

\subsection{Further Applications}

Other applications of compressed sensing include coding and
information theory, machine learning, hyperspectral imaging,
geophysical data analysis, computational biology, remote sensing,
radar analysis, robotics and control, A/D conversion, and many more.
Since an elaborate discussion of all those topics would go beyond
the scope of this survey paper, we refer the interested reader to
\url{dsp.rice.edu/cs}.


\end{document}